\documentclass[12 pt,twoside,aps,prd,amsmath,amssymb,
tightenlines,showpacs,showkeys,eqsecnum]{revtex4}
\usepackage{epsfig}
\usepackage{psfig}
\usepackage{mathptmx}
\usepackage{bm}
\usepackage{graphicx}
\renewcommand{\cal}{\mathcal}

\newcommand {\ve}{\varepsilon}
\newcommand {\cG}{\cal G}
\newcommand {\cD}{\cal D}
\newcommand {\G}{\Gamma}
\newcommand {\bg}{\bar \gamma}
\newcommand {\bp}{\bar \psi}
\newcommand {\bv}{\bar v}
\newcommand {\p}{\psi}

\numberwithin{equation}{section}

\begin{document}
\title{Nonlinear Spinor Field in Cosmology}
\author{Bijan Saha}
\affiliation{Laboratory of Information Technologies\\ 
Joint Institute for Nuclear Research, Dubna\\ 
141980 Dubna, Moscow region, Russia} 
\email{saha@thsun1.jinr.ru}

\begin{abstract}
Within the scope of Bianchi type VI (BVI) model
the self-consistent system of nonlinear spinor and gravitational 
fields is considered. Exact self-consistent solutions to the spinor
and gravitational field equations are obtained for some special choice
of spatial inhomogeneity and nonlinear spinor term. The role of 
inhomogeneity in the evolution of spinor and gravitational field is studied. 
Oscillatory mode of expansion of the BVI universe is obtained for some 
special choice of spinor field nonlinearity. 
\end{abstract}

\maketitle

\bigskip

\section{Introduction}
\setcounter{equation}{0}

The problem of initial singularity still remains at the center
of modern day cosmology. Though Big Bang theory is deep rooted 
among the scientists dealing with early day cosmology, it is 
natural to look back if one can model a Universe free from
initial singularity. An attempt was made by us for several years.
In doing so, we studied the nonlinear spinor field in a 
Bianchi-type I universe (BI) in a series of papers~
\cite{sapfu1,sactp1,sajmp,sactp2,saizv,sagrg,pfu-l,sal,saprd}.

Why nonlinear spinor field? 
It is well-known that
the nonlinear generalization of classical field theory remains
one of the possible ways to over come the difficulties of the 
theory which considers elementary particles as mathematical 
points. In this approach elementary particles are modeled by 
regular (soliton-like) solutions of corresponding nonlinear 
equations. The gravitational field equation is nonlinear by 
nature and the field itself is universal and unscreenable.
These properties lead to definite physical interest for the proper
gravitational field to be considered. We prefer spinor field to
scalar or electromagnetic ones, as the spinor field is the most 
sensitive to the proper gravitational one. 

Why anisotropic universe? 
Though spatially homogeneous and isotropic 
Friedmann-Robertson-Walker (FRW) models are widely considered 
as good approximation of the present and early stages of the 
universe, the large scale matter distribution in the observable 
universe, largely manifested in the form of discrete structures, 
does not exhibit homogeneity of a higher order. In contrast, 
the cosmic background radiation, which is significant in the 
microwave region, is extremely homogeneous, however, recent space
investigations detect anisotropy in the cosmic microwave
background. The observations from cosmic background explorers
differential radiometer have detected and measured cosmic
microwave background anisotropies in different angular scales. 
These anisotropies are supposed to hide in their fold the entire 
history of cosmic evolution dating back to the recombination era
and are being considered as indicative of the geometry and the 
content of the universe. More about cosmic microwave background
anisotropy is expected to be uncovered by the investigations of
microwave anisotropy probe. There is widespread consensus among
the cosmologists that cosmic microwave background anisotropies in
small angular scales have the key to the formation of discrete
structure. The theoretical arguments~\cite{misner} and recent 
experimental data that support the existence of an anisotropic 
phase that approaches an isotropic one leads to consider the
models of universe with anisotropic background. 
A BI universe, being the straightforward generalization of the 
flat FRW universe, is one of the simplest models of an anisotropic 
universe that describes a homogeneous and spatially flat universe.
Unlike the FRW universe, which has the same scale factor for each 
of the three spatial directions, a BI universe has a different 
scale factor in each direction, thereby introducing an
anisotropy to the system. It moreover has the agreeable property 
that near the singularity it behaves like a Kasner universe, even 
in the presence of matter, and consequently falls within the 
general analysis of the singularity given by Belinskii et al.~
\cite{belinskii}. Also in a universe filled with matter for 
$p\,=\,\zeta\,\ve, \quad \zeta < 1$, it has been shown that 
any initial anisotropy in a BI universe quickly dies away and 
a BI universe eventually evolves into a FRW universe~
\cite{jacobs}. Since the present-day universe is surprisingly 
isotropic, this feature of the BI universe 
makes it a prime candidate for studying the possible effects of an 
anisotropy in the early universe on present-day observations. 
In light of the importance mentioned above, several authors have 
studied BI universe from different aspects. In these papers we 
considered the nonlinear spinor field, as well as
a system of interacting spinor and scalar fields. Beside the 
spinor fields, we also study the role of a $\Lambda$ term in the 
evolution of the universe. 

As it is mentioned earlier, in our previous papers we limited our study
within the scope of BI universe for it's simplicity and it's quick transition
to a FRW universe in the process of evolution. But there are few other 
models, which describe an anisotropic space-time and generate particular
interest among physicists~\cite{Singh}. The aim of this report is to study
the self-consistent system of nonlinear spinor and anisotropic
gravitational fields. As an anisotropic space-time we choose Bianchi
type VI (BVI) model, since a suitable choice of it's parameters evokes
few other Bianchi models including BI and FRW universe. It can be noted that
unlike the BI universe, BVI space-time is an inhomogeneous one. Inclusion
of inhomogeneity in the gravitational field significantly complicates the 
search for an exact solution to the system. In the Section 2, we write the
equations for nonlinear spinor fields and the system of Einstein equations.
In this section we also give their solutions in some general form, more 
precisely, we write the solutions in terms of a time-depended function that
can be defined only given the concrete form of the spinor field nonlinearity.
In Section 3, we give exact solutions to the equations for some special
choices of spinor field nonlinearity and space-time inhomogeneity. Beside
this, we also present some numerical solutions in graphical form.

\noindent
\vskip 5mm
\section{Fundamental equations and general solutions}

We shall investigate a self-consistent system of nonlinear
spinor and Einstein gravitational fields. These two fields
are to be codetermined by the following action:
\begin{equation}
{\cal S}({\rm g};\p,\bp) = \int L \sqrt{-{\rm g}}{\rm d \Omega}
\label{action}
\end{equation}
with
\begin{equation}
L=L_{\rm grav.} + L_{\rm spinor}.
\label{glag}
\end{equation}
The gravitational part of the Lagrangian \eqref{glag} is given
by a Bianchi type-VI (BVI) space-time, while the spinor part
is a usual spinor field Lagrangian with an arbitrary nonlinear
term.
  
\subsection{Spinor field Lagrangian}

For a spinor field $\p$, symmetry between $\p$ and $\bp$ appears to
demand that one should choose the symmetrized Lagrangian~
\cite{kibble}.
Keep it in mind we choose the spinor field Lagrangian with a 
nonlinear term in \eqref{lag} as follows:
\begin{equation} 
L_{\rm spinor}=\frac{i}{2} 
\biggl[ \bp \gamma^{\mu} \nabla_{\mu} \p- \nabla_{\mu} \bp 
\gamma^{\mu} \p \biggr] - M\bp \p + L_N.  \label{lag}
\end{equation} 
Here $M$ is the spinor mass.The nonlinear term $L_N$ describes 
the self-interaction of spinor field and can be presented
as some arbitrary functions of invariants generated from the 
real bilinear forms of spinor field having the form: 
\begin{subequations}
\label{bf}
\begin{eqnarray}
 S&=& \bar \psi \psi\qquad ({\rm scalar}),   \\                   
  P&=& i \bar \psi \gamma^5 \psi\qquad ({\rm pseudoscalar}), \\
 v^\mu &=& (\bar \psi \gamma^\mu \psi) \qquad ({\rm vector}),\\
 A^\mu &=&(\bar \psi \gamma^5 \gamma^\mu \psi)\qquad 
({\rm pseudovector}), \\
Q^{\mu\nu} &=&(\bar \psi \sigma^{\mu\nu} \psi)\qquad
({\rm antisymmetric\,\,\, tensor}),  
\end{eqnarray}
\end{subequations}
where $\sigma^{\mu\nu}\,=\,(i/2)[\gamma^\mu\gamma^\nu\,-\,
\gamma^\nu\gamma^\mu]$. 
Invariants, corresponding to the bilinear forms, are
\begin{subequations}
\label{invariants}
\begin{eqnarray}
I &=& S^2, \\
J &=& P^2, \\ 
I_v &=& v_\mu\,v^\mu\,=\,(\bar \psi \gamma^\mu \psi)\,g_{\mu\nu}
(\bar \psi \gamma^\nu \psi),\\ 
I_A &=& A_\mu\,A^\mu\,=\,(\bar \psi \gamma^5 \gamma^\mu \psi)\,
g_{\mu\nu}(\bar \psi \gamma^5 \gamma^\nu \psi), \\
I_Q &=& Q_{\mu\nu}\,Q^{\mu\nu}\,=\,(\bar \psi 
\sigma^{\mu\nu} \psi)\,g_{\mu\alpha}g_{\nu\beta}
(\bar \psi \sigma^{\alpha\beta} \psi). 
\end{eqnarray}
\end{subequations}

According to the Pauli-Fierz theorem \cite{Ber} among the five 
invariants only $I$ and $J$ are independent as all other can be 
expressed by them:
$I_V = - I_A = I + J$ and $I_Q = I - J.$ Therefore, we choose 
the nonlinear term $L_N = F(I, J)$, thus claiming that it 
describes the nonlinearity in the most general of its form. 

\subsection{The gravitational field}

The gravitational part of the Lagrangian in \eqref{glag} has the 
form:
\begin{equation}
L_{\rm grav.} = \frac{R}{2 \kappa},
\label{gr}
\end{equation}
Here $R$ is the scalar curvature and $\kappa$ is  the  
Einstein's gravitational constant. The gravitational 
field in our case is given by a Bianchi-type VI (BVI) metric:
\begin{equation} 
ds^2 = dt^2 - a^{2} e^{-2mz}\,dx^{2} - b^{2} e^{2nz}\,dy^{2} - 
c^{2}\,dz^2, 
\label{bvi}
\end{equation}
with $a,\,b,\,c$ being the functions of time only. Here 
$m,\,n$ are some arbitrary constants and the velocity of light is 
taken to be unity. It should be emphasized that the BVI metric
models a Universe that is anisotropic and inhomogeneous. A 
suitable choice of $m,\,n$ as well as the metric functions 
$a,\,b,\,c$ in the BVI given by \eqref{bvi} 
evokes the following Bianchi-type universes. Thus
\begin{itemize}
\item
for $m=n$ the BVI metric transforms to a Bianchi-type V (BV) one, 
i.e.,
$m = n$, BVI $\Longrightarrow$ BV $\in$ open FRW with the line 
elements
\begin{equation} 
ds^2 = dt^2 - a^{2} e^{-2mz}\,dx^{2} - b^{2} e^{2mz}\,dy^{2} - 
c^{2}\,dz^2; 
\label{bv}
\end{equation}
\item
for $n=0$ the BVI metric transforms to a Bianchi-type III (BIII) 
one, i.e.,
$n = 0$, BVI $\Longrightarrow$ BIII with the line elements
\begin{equation} 
ds^2 = dt^2 - a^{2} e^{-2mz}\,dx^{2} - b^{2} \,dy^{2} - 
c^{2}\,dz^2; 
\label{biii}
\end{equation}
\item
for $m=n =0$ the BVI metric transforms to a Bianchi-type I (BI) 
one, i.e.,
$m = n = 0$, BVI $\Longrightarrow$ BI with the line elements
\begin{equation} 
ds^2 = dt^2 - a^{2} \,dx^{2} - b^{2} \,dy^{2} - c^{2}\,dz^2; 
\label{bi}
\end{equation}
\item
for $m=n=0$ and equal scale factor in all three directions the 
BVI metric 
transforms to a Friedmann-Robertson-Walker (FRW) universe, i.e.,
$m = n = 0$ and $a=b=c$, BVI $\Longrightarrow$ FRW with the line 
elements.
\begin{equation} 
ds^2 = dt^2 - a^{2} \bigl(dx^{2} + \,dy^{2} + \,dz^2\bigr). 
\label{frw}
\end{equation}
\end{itemize}

Let us write the nontrivial components of Ricci and Riemann tensors
as well as Christoffel symbols of the BVI metric.

The nontrivial Christoffel symbols
of the BVI metric read
\begin{eqnarray}
\G_{01}^{1} &=& \dot{a}/a,\quad
\G_{02}^{2} = \dot{b}/b,\quad
\G_{03}^{3} = \dot{c}/c, \nonumber\\
\G_{11}^{0} &=& a \dot{a} e^{-2mz},\quad
\G_{22}^{0} = b \dot{b} e^{2nz},\quad
\G_{33}^{0} = c \dot{c},\nonumber\\
\G_{31}^{1} &=& -m,\quad
\G_{32}^{2} = n,\quad
\G_{11}^{3} = \frac{m a^2}{c^2} e^{-2mz},\quad
\G_{22}^{3} = -\frac{n b^2}{c^2} e^{2mz}. \nonumber
\end{eqnarray} 
The nontrivial components of Riemann tensor are
\begin{eqnarray}
R_{\,\,\,01}^{01} &=& -\frac{\ddot a}{a}, \quad
R_{\,\,\,02}^{02} = -\frac{\ddot b}{b}, \quad
R_{\,\,\,03}^{03} = -\frac{\ddot c}{c}, \nonumber \\
R_{\,\,\,12}^{12} &=& -\frac{mn}{c^2} -
\frac{\dot a}{a}\frac{\dot b}{b},\quad
R_{\,\,\,13}^{13} = \frac{m^2}{c^2} -
\frac{\dot c}{c}\frac{\dot a}{a}, \quad
R_{\,\,\,23}^{23} = \frac{n^2}{c^2} -
\frac{\dot b}{b}\frac{\dot c}{c}.
\nonumber
\end{eqnarray}

The nontrivial components of the Ricci tensor are
\begin{eqnarray}
R_{3}^{0} &=& \Bigl(m\frac{\dot a}{a} - n\frac{\dot b}{b} -
(m -n ) \frac{\dot c}{c}\Bigr), \nonumber\\
R_{0}^{0} &=& -\Bigl(\frac{\ddot a}{a^2} + \frac{\ddot b}{b^2} + 
\frac{\ddot c}{c^2}\Bigr), \nonumber\\
R_{1}^{1} &=& - \Bigl(\frac{\ddot a}{a}+ 
\frac{\dot a}{a}\frac{\dot b}{b} + \frac{\dot a}{a}\frac{\dot c}{c}
-\frac{m^2 - mn}{c^2}\Bigr), \nonumber\\
R_{2}^{2} &=& - \Bigl(\frac{\ddot b}{b}+ 
\frac{\dot a}{a}\frac{\dot b}{b} + \frac{\dot b}{b}\frac{\dot c}{c}
-\frac{n^2 - mn}{c^2}\Bigr), \nonumber\\
R_{3}^{3} &=& - \Bigl(\frac{\ddot c}{c}+ 
\frac{\dot a}{a}\frac{\dot c}{c} + \frac{\dot b}{b}\frac{\dot c}{c}
-\frac{m^2 + n^2}{c^2}\Bigr), \nonumber
\end{eqnarray}
To investigate the existence of singularity (singular point), one
has to study the invariant characteristics of space-time.
As we know, in general relativity, these invariants are composed
from the curvature tensor and the metric one. Though, in 4D
Riemann space there are 14 independent invariants, it is 
sufficient to study only three of them, namely the scalar
curvature $I_1 = R$, $I_2 = R_{\mu \nu}R^{\mu\nu}$, and the
Kretschmann scalar 
$I_3 = R_{\alpha \beta \mu \nu} R^{\alpha \beta \mu \nu}$.
From the Riemann and Ricci tensors written above one can easily
write find
\begin{equation}
I_1=R = -\frac{2}{\tau}\bigl[{\ddot \tau}-{\dot a}{\dot b}c
-a {\dot b} {\dot c}-{\dot a}b{\dot c}-ab(m^2 - mn + n^2)/c 
\bigr], \qquad \tau = a b c,
\end{equation}                                          
whereas $I_2 \propto 1/(\tau c)^2$ and $I_3 \propto 1/(\tau c)^2$.
Thus we see that at any space-time point, where $\tau = 0$
the invariants $I_1, I_2, I_3$ become infinity, hence the 
space-time becomes singular at this point.

\subsection{Field equations}
The field equations for the spinor and gravitational fields
can be obtained from variational principle.
Variation of the Lagrangian \eqref{lag} with respect to 
field functions $\psi (\bp)$  gives the nonlinear spinor field 
equations:
\begin{subequations}
\label{spin}
\begin{eqnarray}
i\gamma^\mu \nabla_\mu \p -M\p + {\cD} \p + i {\cG} 
\gamma^5 \p &=& 0 \\
i \nabla_\mu \bp \gamma^\mu + M \bar\psi - {\cD} \bp - 
i {\cG} \bp \gamma^5 &=& 0 
\end{eqnarray}
\end{subequations}
where ${\cD} = 2 S F_I$ and ${\cG} = 2 P F_J$.  

Varying \eqref{lag} with respect to metric function ($g_{\mu\nu}$)
we find the Einstein's field equation 
\begin{equation}
R_{\nu}^{\mu} - \frac{1}{2}\delta_{\nu}^{\mu} R =
\kappa T_{\nu}^{\mu},
\label{eing}
\end{equation}
where $R_{\nu}^{\mu}$ is the Ricci tensor, $R$ is the Ricci scalar,
and $T_{\nu}^{\mu}$ is the energy-momentum tensor of the spinor
field. In our case, where space-time is given by a BVI metric
\eqref{bvi}, the equations for the metric functions $a,\,b,\,c$ 
read  
\begin{subequations}
\label{ein}
\begin{eqnarray}
\frac{\ddot b}{b} +\frac{\ddot c}{c} +\frac{\dot b}{b}\frac{\dot 
c}{c} - \frac{n^2}{c^2} &=& \kappa T_{1}^{1}, \label{11}\\
\frac{\ddot c}{c} +\frac{\ddot a}{a} +\frac{\dot c}{c}\frac{\dot 
a}{a} - \frac{m^2}{c^2} &=& \kappa T_{2}^{2}, \label{22} \\
\frac{\ddot a}{a} +\frac{\ddot b}{b} +\frac{\dot a}{a}\frac{\dot 
b}{b} + \frac{m n}{c^2} &=& \kappa T_{3}^{3}, \label{33}\\
\frac{\dot a}{a}\frac{\dot b}{b} +\frac{\dot b}{b}
\frac{\dot c}{c} +
\frac{\dot c}{c}\frac{\dot a}{a} - \frac{m^2 - m n + n^2}{c^2} &=& 
\kappa T_{0}^{0}, \label{00}\\
m \frac{\dot a}{a} - n \frac{\dot b}{b}
- (m - n) \frac{\dot c}{c} &=& \kappa T_{3}^{0}. \label{03}
\end{eqnarray}
\end{subequations}
Here over dots denote differentiation with respect to time ($t$). 
The energy-momentum tensor of the material field $T_{\mu}^{\nu}$ 
has the form:
\begin{equation}
T_{\mu}^{\rho}=\frac{i}{4} g^{\rho\nu} \biggl(\bp \gamma_\mu 
\nabla_\nu \psi + \bp \gamma_\nu \nabla_\mu \psi - \nabla_\mu \bp 
\gamma_\nu \psi - \nabla_\nu \bp \gamma_\mu \psi \biggr) \,-
\delta_{\mu}^{\rho}L_{sp}.
\label{emt}
\end{equation}
Here $L_{sp}$ is the spinor field Lagrangian, which 
on account of spinor field Eqns.~ \eqref{spin} takes the form:
\begin{equation}
L_{sp} = - {\cD} S - {\cG P} + F.
\label{sflag}
\end{equation}
In the expressions above $\nabla_\mu$ denotes the covariant 
derivative of spinor, having the form \cite{Zelnor}:  
\begin{equation} \nabla_\mu 
\p=\partial_\mu \p - \Gamma_\mu \p 
\end{equation} 
where $\Gamma_\mu$ is spin connection. The spin affine 
connection matrices $\Gamma_\mu (x)$ are uniquely determined
up to an additive multiple of the unit matrix by the 
Eqn.~\cite{brill}
\begin{equation}
\partial_{\mu} \gamma_{\nu} - \Gamma^{\alpha}_{\nu\mu}
\gamma_{\alpha} - \Gamma_{\mu}\gamma_{\nu}
+ \gamma_{\nu}\Gamma_{\mu}=0,
\label{spcon} 
\end{equation}
with the solution
\begin{equation}
\Gamma_\mu (x)= \frac{1}{4}g_{\rho \sigma}(x)
\bigl(\partial_{\mu}e_{\delta}^{b} e_{b}^{\rho} - 
\Gamma_{\mu \delta}^{\rho}\bigr)\gamma^{\sigma}\gamma^{\delta}.
\end{equation}
For the metric element \eqref{bvi} it gives
\begin{eqnarray} 
\Gamma_0 &=& 0, \nonumber\\ 
\Gamma_1 &=& \frac{1}{2}\bigl[{\dot a} \bg^{1} \bg^0
- m \frac{a}{c} \bg^{1} \bg^3 \bigr] e^{-mz} \nonumber \\  
\Gamma_2 &=& \frac{1}{2}\bigl[ {\dot b} \bg^{2} \bg^0
+ n \frac{b}{c} \bg^{2} \bg^3 \bigr] e^{nz}, \nonumber\\
\Gamma_3 &=& \frac{1}{2} {\dot c} \bg^{3} \bg^0  \nonumber
\end{eqnarray}
It is easy to show that
$$\gamma^\mu \Gamma_\mu = -\frac{1}{2}\frac{\dot \tau}{\tau}\bg^0
+ \frac{m - n}{2 c}\bg^3,$$
where we define 
\begin{equation}
\tau = a b c.
\label{taudef}
\end{equation}
The Dirac matrices $\gamma^\mu(x)$ of curved space-time are 
connected with those of Minkowski one as follows:
$$ \gamma^0=\bg^0,\quad \gamma^1 =\bg^1 e^{mz}/a,
\quad \gamma^2=\bg^2 /b e^{nz},\quad \gamma^3 =\bg^3 /c$$
with 
\begin{eqnarray}
\bg^0\,=\,\left(\begin{array}{cc}I&0\\0&-I\end{array}\right), \quad
\bg^i\,=\,\left(\begin{array}{cc}0&\sigma^i\\
-\sigma^i&0\end{array}\right), \quad
\gamma^5 = \bg^5&=&\left(\begin{array}{cc}0&-I\\
-I&0\end{array}\right),\nonumber
\end{eqnarray}
where $\sigma_i$ are the Pauli matrices:
\begin{eqnarray}
\sigma^1\,=\,\left(\begin{array}{cc}0&1\\1&0\end{array}\right), 
\quad
\sigma^2\,=\,\left(\begin{array}{cc}0&-i\\i&0\end{array}\right), 
\quad
\sigma^3\,=\,\left(\begin{array}{cc}1&0\\0&-1\end{array}\right).
\nonumber
\end{eqnarray}
Note that the $\bg$ and the $\sigma$ matrices obey the following 
properties:
\begin{eqnarray}
\bg^i \bg^j + \bg^j \bg^i = 2 \eta^{ij},\quad i,j = 0,1,2,3 
\nonumber\\
\bg^i \bg^5 + \bg^5 \bg^i = 0, \quad (\bg^5)^2 = I, 
\quad i=0,1,2,3 \nonumber\\
\sigma^j \sigma^k = \delta_{jk} + i \varepsilon_{jkl} \sigma^l, 
\quad j,k,l = 1,2,3 \nonumber
\end{eqnarray}
where $\eta_{ij} = \{1,-1,-1,-1\}$ is the diagonal matrix, 
$\delta_{jk}$ is the Kronekar symbol and $\varepsilon_{jkl}$ 
is the totally antisymmetric matrix with $\varepsilon_{123} = +1$.
Let us consider the spinors to be functions of $t$ and $z$ only, 
such that
\begin{equation}
\p (t, z) = v (t) e^{i k z}, \quad \bp (t,z) = \bv (t) e^{- i k z}
\label{z}
\end{equation}

Inserting \eqref{z} into \eqref{spin} for the nonlinear spinor 
field we find
\begin{subequations}
\label{spinv}
\begin{eqnarray}
\bg^0\Bigl(\dot v + \frac{\dot \tau}{2 \tau} v\Bigr) 
-\Bigl(\frac{m-n}{2 c} - i\frac{k}{c}\Bigr)\bg^3 v + i\Phi v 
+ {\cG} \bg^5 v &=& 0 \\
\Bigl(\dot{\bv} + \frac{\dot \tau}{2 \tau}\bv\Bigr)\bg^0 
-\Bigl(\frac{m-n}{2 c} + i\frac{k}{c}\Bigr) \bv \bg^3 - i\Phi \bv  
- {\cG} \bv \bg^5  &=& 0. 
\end{eqnarray}
\end{subequations}
Here we define $\Phi = M - {\cD}$.
Let us introduce a new function 
$$u_j (t) = \sqrt{\tau} v_j(t).$$ 
Then for the components of the NLSF from \eqref{spinv} one obtains
\begin{subequations}
\label{u}
\begin{eqnarray} 
\dot{u}_{1} + i \Phi u_{1} - 
\Bigl[\frac{m-n}{2 c} - i\frac{k}{c} +{\cG}\Bigr] u_{3} &=& 0, \\
\dot{u}_{2} + i \Phi u_{2} + 
\Bigl[\frac{m-n}{2 c} - i\frac{k}{c} -{\cG}\Bigr] u_{4} &=& 0, \\
\dot{u}_{3} - i \Phi u_{3} - 
\Bigl[\frac{m-n}{2 c} - i\frac{k}{c} -{\cG}\Bigr] u_{1} &=& 0, \\
\dot{u}_{4} - i \Phi u_{4} + 
\Bigl[\frac{m-n}{2 c} - i\frac{k}{c} +{\cG}\Bigr] u_{2} &=& 0. 
\end{eqnarray} 
\end{subequations}

Using the spinor field Eqns. \eqref{spin} and \eqref{spinv}
it can be shown that the bilinear spinor forms, defined
by \eqref{bf}, i.e.,  
\begin{eqnarray}
S &=& \bp \p = \bv v, \quad            
P = i \bp \bg^5 \p = i \bv \bg^5 v, \quad    
A^{0} = \bp \bg^5 \bg^0 \p = \bv \bg^5 \bg^0 v, \nonumber \\
A^{3} &=& \bp \bg^5 \bg^3 \p = \bv \bg^5 \bg^3 v, \quad
V^{0} = \bp \bg^0 \p = \bv \bg^0 v, \quad
V^{3} = \bp \bg^3 \p = \bv \bg^3 v, \nonumber \\
Q^{30} &=& i \bp \bg^3 \bg^0 \p = i \bv \bg^3 \bg^0 v, \quad
Q^{21} = \bp  \bg^0 \bg^3 \bg^5 \p = 
i \bp  \bg^2 \bg^1 \p = i \bv  \bg^2 \bg^1 v, \nonumber 
\end{eqnarray}
obey the following system of equations: 
\begin{subequations}
\label{inv}
\begin{eqnarray}                                
\dot S_0 -2 \frac{k}{c} Q_{0}^{30} - 2 {\cG} A_{0}^{0} &=& 0, \\
\dot P_0 -2 \frac{k}{c} Q_{0}^{21} - 2 \Phi A_{0}^{0} &=& 0, \\
\dot A_{0}^{0} -\frac{m-n}{c} A_{0}^{3} + 2 \Phi P_0 + 2 {\cG} 
S_0 &=& 0,\\
\dot A_{0}^{3} -\frac{m-n}{c} A_{0}^{0} &=& 0, \\                 
\dot V_{0}^{0} - \frac{m-n}{c} V_{0}^{3} &=& 0, \\                
\dot V_{0}^{3} - \frac{m-n}{c} V_{0}^{0} + 
2 \Phi Q_{0}^{30} - 2 {\cG} Q_{0}^{21} &=& 0,\\
\dot Q_{0}^{30} + 2\frac{k}{c}S_0 - 2 \Phi V_{0}^{3} &=& 0, \\    
\dot Q_{0}^{21} +2 \frac{k}{c}P_0 + 2 {\cG} V_{0}^{3} &=& 0,
\end{eqnarray}
\end{subequations}
where we denote $F_0 = \tau F$. Combining these equations 
together and taking the first integral one gets
\begin{equation}                                
(S_{0})^{2} + (P_{0})^{2} + (A_{0}^{0})^{2} - (A_{0}^{3})^{2} -
(V_{0}^{0})^{2} + (V_{0}^{3})^{2} + 
(Q_{0}^{30})^{2} + (Q_{0}^{21})^{2} = C = {\rm Const} 
\label{I2}             
\end{equation}

Before dealing with the Einstein Eqns. \eqref{ein} let us go 
back to \eqref{u}. From the first and the third equations of the 
system \eqref{u} one finds
\begin{equation}
\dot{u}_{13} = ({\cG} - Q) u_{13}^{2} - 2 i \Phi u_{13}
+ ({\cG} + Q),
\label{rik}
\end{equation}
where, we denote $u_{13} = u_1/u_3$ and $Q=[m - n - 2ik]/2c$.
The Eqn. \eqref{rik} is a Riccati one~\cite{Kamke} with 
variable coefficients. A transformation~\cite{Zaitsev}  
\begin{equation}
v_{13} = {\rm exp}\Bigl(- \int ({\cG} - Q) u_{13} dt\Bigr), 
\end{equation}
leads the general Riccati Eqn. \eqref{rik} to the second order linear 
one, namely,
\begin{equation}
({\cG} -Q) \ddot{v}_{13} + \bigl[ 2 i \Phi ({\cG} - Q) - 
\dot{{\cG}} + \dot{Q}\bigr]\dot{v}_{13} + ({\cG} - Q)^2
({\cG} + Q) v_{13} = 0.
\label{v13}
\end{equation}        
Some times it is easier to solve a linear second order differential 
equation than a first order nonlinear one. But we also give a general 
solution to \eqref{rik}. For this purpose we rewrite \eqref{rik} in the form
\begin{equation}
\dot{w}_{13} = ({\cG} - Q) w_{13}^{2} e^{-2i \int \Phi(t) dt}
+ ({\cG} + Q)e^{2i \int \Phi(t) dt},
\label{w13}
\end{equation}  
where we set $u_{13} = w_{13} {\rm exp}[-2i \int \Phi(t) dt].$
It is an inhomogeneous nonlinear differential equation. The solution for the
homogeneous part of \eqref{w13}, i.e.,
\begin{equation}
\dot{w}_{13} = ({\cG} - Q) w_{13}^{2} e^{-2i \int \Phi(t) dt}
\label{w13h}
\end{equation}
reads
\begin{equation}
w_{13} = -\Biggl[\int ({\cG} - Q) e^{-2i \int \Phi(t) dt} dt + C
\Biggr]^{-1},
\label{solh}
\end{equation}  
with $C$ being some arbitrary constant. Then the general solution to the 
inhomogeneous Eqn. \eqref{w13} can be presented as
\begin{equation}
w_{13} = -\Biggl[\int ({\cG} - Q) e^{-2i \int \Phi(t) dt} dt + C(t)
\Biggr]^{-1},
\label{solgen}
\end{equation}
with time dependent parameter $C(t)$ to be determined from
\begin{equation}
\dot{C} =
\Biggl[\int ({\cG} - Q) e^{-2i \int \Phi(t) dt} dt + C(t)
\Biggr]^{2} ({\cG} + Q) e^{2i \int \Phi(t) dt}.
\label{Cd}
\end{equation} 

Thus given the concrete nonlinear term in the Lagrangian and solution
of the Einstein equations, one finds the relation between 
$u_1$ and $u_3$ ($u_2$ and $u_4$ as well), hence the components
of the spinor field. 

Now we study the Einstein Eqns. \eqref{ein}. In doing so,
we write the components of the energy-momentum tensor, which 
in our case read
\begin{subequations}
\label{cemt}
\begin{eqnarray}
T_{0}^{0}&=& mS - F + \frac{k}{c} V^3, \\
T_{1}^{1} &=& T_{2}^{2} = {\cD} S + {\cG} P - F, \\  
T_{3}^{3}&=& {\cD} S + {\cG} P - F - \frac{k}{c} V^3, \\
T_{3}^{0} &=& - k V^0.  
\end{eqnarray}
\end{subequations}
Let us demand the energy-momentum tensor to be conserved, i.e.,
\begin{equation}
T_{\nu;\mu}^{\mu} = T_{\nu,\mu}^{\mu} + \Gamma_{\beta \mu}^{\mu}\,
T_{\nu}^{\beta} - \Gamma_{\nu \mu}^{\beta}\,T_{\beta}^{\mu} = 0
\label{emc}
\end{equation}
Taking into account that $T_{\mu}^{\nu}$ is a function of $t$ 
only, from \eqref{emc} we find
\begin{subequations}
\label{emcex}
\begin{eqnarray}
\Phi \dot S_{0} - {\cG} \dot P_{0} +\frac{k}{c} \dot V_{0}^{3}
- \frac{k}{c} \frac{m - n}{c} V_{0}^{0} &=& 0, \\
\dot V_{0}^{0} - \frac{m - n}{c} V_{0}^{3} &=& 0.
\end{eqnarray}
\end{subequations}
As one can easily verify, the Eqns. \eqref{emcex} are 
consistent with those of \eqref{inv}. 

Let us go back to the Eqns. \eqref{ein}.
In view of \eqref{cemt}, from \eqref{03} one obtains 
\begin{equation}
\Bigl(\frac{a}{c}\Bigr)^m = \Bigl(\frac{b}{c}\Bigr)^n
 {\cal N} {\rm exp}[-\kappa k \int V^0 dt], 
\quad {\cal N} = {\rm const.}.
\label{abcrel}
\end{equation}
Subtracting \eqref{11} from \eqref{22} we find 
\begin{equation}
\frac{d}{d t} 
\Bigl[\tau \frac{d}{dt}\Bigl\{{\rm ln}\Bigl(\frac{a}{b}\Bigr)
\Bigr\}\Bigr] = \frac{m^2 - n^2}{c^2} \tau.
\label{ab1}
\end{equation}
Analogically, subtraction of \eqref{11} from \eqref{33} and
\eqref{22} from \eqref{33} give
\begin{equation}
\frac{d}{d t} 
\Bigl[\tau \frac{d}{dt}\Bigl\{{\rm ln}\Bigl(\frac{a}{c}\Bigr)
\Bigr\}\Bigr] = -\frac{mn + n^2}{c^2} \tau - \frac{\kappa k}{c} V^3 \tau,
\label{ac1}
\end{equation}
and
\begin{equation}
\frac{d}{d t} 
\Bigl[\tau \frac{d}{dt}\Bigl\{{\rm ln}\Bigl(\frac{b}{c}\Bigr)
\Bigr\}\Bigr] = -\frac{mn + m^2}{c^2} \tau - \frac{\kappa k}{c} V^3 \tau,
\label{bc1}
\end{equation}
respectively. It can be shown that, in view of \eqref{inv} and 
\eqref{abcrel} the Eqns. \eqref{ab1}, \eqref{ac1} and \eqref{bc1} are
interchangeable. Thus in \eqref{abcrel} we have inter relation between
the metric functions, whereas two more equations for defining any two 
of the three.
Taking into account that $\tau = a b c$, from \eqref{abcrel} we can write
$a$ and $b$ in terms of $c$, such that
\begin{equation}
a = \Bigl[\tau^n c^{m-2n} {\cal N} {\rm exp}[-\kappa k \int V^0 dt]
\Bigr]^{1/(m+n)},
\label{atoc}
\end{equation} 
and 
\begin{equation}
b = \Bigl[\tau^m c^{n-2m}/\bigl({\cal N} {\rm exp}[-\kappa k \int V^0 dt]
\bigr)\Bigr]^{1/(m+n)}. 
\label{btoc}
\end{equation}
In view of \eqref{atoc}, \eqref{btoc} and \eqref{inv} from \eqref{ab1} 
one finds
\begin{equation}
\frac{\ddot \tau}{\tau} = 3 \frac{\dot \tau}{\tau} \frac{\dot c}{c} +
3\Bigl(\frac{\ddot c}{c} - \frac{{\dot c}^2}{c^2}
\Bigr) - 2 \frac{\kappa k}{c} V^3 - \frac{(m+n)^2}{c^2}.
\label{tau3}
\end{equation}
In getting \eqref{tau3} we employ only four out of five Einstein equations,
leaving \eqref{00} unused. 

On the other hand, adding \eqref{11}, \eqref{22}, \eqref{33} and \eqref{00}, 
multiplied by 3 we get the equation for $\tau$, which in view of \eqref{cemt}
takes the form
\begin{equation}
\frac{\ddot \tau}{\tau} = 2 \frac{m^2 - mn + n^2}{c^2} + 
\frac{\kappa}{2} \bigl[3(MS + {\cD} S + {\cG} P - 2 F) + 
2 \frac{k}{c} V^3\bigr].
\label{detertau}
\end{equation}
Thus we are left with two equations, namely \eqref{tau3} and 
\eqref{detertau} for two unknowns $c$ and $\tau$. These two equations can be 
combined to get
\begin{equation}
\frac{\ddot c}{c} - \frac{{\dot c}^2}{c^2}+ 
\frac{\dot \tau}{\tau} \frac{\dot c}{c} =\frac{\kappa k}{c} V^3
+ \frac{m^2 + n^2}{c^2} +
\frac{\kappa}{2} \bigl[MS + {\cD} S + {\cG} P - 2 F\bigr].
\label{ctau}
\end{equation}
Thus we come to the Eqn. \eqref{ctau} where all the equations in hand,
both spinor and gravitational, are employed. Moreover, both $c$ and $\tau$
are inter-related. Assuming $c$ as a function of $\tau$ (vice versa)
and giving the concrete form of spinor field nonlinearity one finds the 
solution of \eqref{ctau} exactly what we do in the section to follow.

\section{Analysis of the result}

In the previous section we derived the fundamental equations for nonlinear 
spinor fields and metric functions. Comparing the equation with those of 
in a BI universe (see e.g., \cite{saprd}) we conclude that introduction of 
inhomogeneity both in gravitational (through $m$ and $n$) and spinor 
(through $k$) fields significantly complicates the whole picture. 
In what follows, we try to write the solutions in a more explicit form
for some special choice of spinor field nonlinearity and field and 
space-time inhomogeneity. 

Let us first consider the Eqn. \eqref{ctau}. As one sees, there are two
unknown functions in this equation, namely, $c$ and $\tau$. Though, they are
inter-related, there is no equation giving this relation explicitly. So, as 
a first step, we have to make some assumption relating $c$ and $\tau$. In what
follows, we consider few models giving explicit relation between $c$ and 
$\tau$ and study them for different types of nonlinear spinor terms.

{\bf Case I.} Let us assume that
\begin{equation}           
c = \tau.
\label{tc}
\end{equation}
Under this assumption in view of $\tau = a b c$, we should have $a =1/b$.
Indeed, from \eqref{atoc} and \eqref{btoc} we find
\begin{equation}
a = \Bigl[\tau^{m-n} {\cal N} {\rm exp}[-\kappa k \int V^0 dt]
\Bigr]^{1/(m+n)},
\label{atoctau}
\end{equation} 
and 
\begin{equation}
b = \Bigl[\tau^{n-m}/\bigl({\cal N} {\rm exp}[-\kappa k \int V^0 dt]
\bigr)\Bigr]^{1/(m+n)}. 
\label{btoctau}
\end{equation}
With regard to \eqref{tc}, from \eqref{ctau} we obtain
\begin{equation}
\frac{\ddot \tau}{\tau} =  \frac{\kappa k}{\tau} V^3 +
 \frac{m^2 + n^2}{\tau^2} +
\frac{\kappa}{2} \bigl[MS + {\cD} S + {\cG} P - 2 F\bigr].
\label{cetau}
\end{equation}
Let us now study \eqref{cetau} for some special choice of spinor field 
nonlinearity. 

\underline{Linear spinor field}

To begin with we consider the linear case setting $F(I,J) = 0$. It 
immediately leads to ${\cal D} = 0$ and ${\cal G} = 0.$ Eqn.
\eqref{cetau} now takes the form
\begin{equation}
\frac{\ddot \tau}{\tau} =  \frac{\kappa k}{\tau} V^3 +
 \frac{m^2 + n^2}{\tau^2} +
\frac{\kappa}{2} MS.
\label{cetaulin}
\end{equation}
As one sees, to solve \eqref{cetaulin}, we have to find $V^3$ and $S$ first. 
From \eqref{inv} for the linear spinor field we have
\begin{subequations}
\label{invlin}
\begin{eqnarray}                                
\dot S_0 -2 \frac{k}{c} Q_{0}^{30} &=& 0, \\
\dot P_0 -2 \frac{k}{c} Q_{0}^{21} - 2 M A_{0}^{0} &=& 0, \\
\dot A_{0}^{0} -\frac{m-n}{c} A_{0}^{3} + 2 M P_0  &=& 0,\\
\dot A_{0}^{3} -\frac{m-n}{c} A_{0}^{0} &=& 0, \\                 
\dot V_{0}^{0} - \frac{m-n}{c} V_{0}^{3} &=& 0, \\                
\dot V_{0}^{3} - \frac{m-n}{c} V_{0}^{0} + 2 M Q_{0}^{30} &=& 0,\\
\dot Q_{0}^{30} + 2\frac{k}{c}S_0 - 2 M V_{0}^{3} &=& 0, \\    
\dot Q_{0}^{21} +2 \frac{k}{c}P_0 &=& 0,
\end{eqnarray}
\end{subequations}
with the first integrals
\begin{subequations}
\label{fintlin}
\begin{eqnarray}                                
(S_{0})^{2} + (V_{0}^{3})^{2} + (Q_{0}^{30})^{2} - (V_{0}^{0})^{2} &=& 0,\\
(P_{0})^{2} + (A_{0}^{0})^{2} + (Q_{0}^{21})^{2} - (A_{0}^{3})^{2} &=& 0.
\end{eqnarray}
\end{subequations}
Thus we see that even in case of linear spinor field with $k \ne 0$ we cannot
write $V^3$ or $S$ explicitly. In order to express $S$ or $P$, hence the
massive term or spinor field nonlinearity, in terms of $\tau$, we now
consider the spinor field to be to be space independent setting $k = 0.$

From \eqref{inv} in this case one obtains,
\begin{equation}
S = \frac{C_0}{\tau}, 
\end{equation}
with $C_0$ being the integration constant.

The Eqn. \eqref{cetau} in this case takes the form
\begin{equation}
\ddot \tau = \frac{m^2 + n^2}{\tau} + \frac{\kappa}{2} M C_0.
\label{c=taulin}
\end{equation}
The solution of \eqref{c=taulin} can be written in quadrature as
\begin{equation}
\int \frac{d \tau}{\sqrt{2(m^2 +n^2)\, {\rm ln}\tau + 
\kappa M C_0\, \tau + E}} = t, \quad E = {\rm const.}
\label{linsolkz}
\end{equation}
For the solution to be meaningful, the integrand in \eqref{linsolkz} should 
be positive. It means that for the $\tau$ to have an initial value close to 
zero, one has to set small value for $m$ and $n$, while the constant $E$
should be large enough.  

Components of spinor field can be obtained from \eqref{solh}. In the case 
considered, ${\cG} = 0$, $Q = (m-n)/2\tau$ and $\Phi = M$.

\underline{Nonlinear spinor field with $k = 0.$}

Let us now consider the nonlinear spatially independent spinor field. 
We first choose the nonlinear term being a function of $I=S^2$ only,
followed by a massless spinor field with the nonlinear term to be a
function of $J=P^2$. 

If the nonlinear spinor term is given as $F = F(I) = \lambda S^\eta$, 
where $\lambda$ is the (self)coupling constant, then in view of
$S = C_0/\tau$ for $\tau$ we find
\begin{equation}
{\ddot \tau} = \frac{m^2 + n^2}{\tau} +
\frac{\kappa}{2} M C_0 + \frac{\kappa \lambda(\eta -2)C_0^\eta}
{2 \tau^{\eta-1}},
\label{cntau}
\end{equation}
with the solution in quadrature
\begin{equation}
\int \frac{d \tau}{\sqrt{2(m^2 +n^2) {\rm ln}\,\tau + \kappa M C_0\,\tau
-\kappa \lambda C_0^\eta\,\tau^{2 - \eta}  
+ E}} = t.
\label{nonlinsolkz}
\end{equation}
As one sees, inclusion of nonlinear term sets additional restriction on
the smallness of the initial value of $\tau$. 
Components of spinor field, as in linear case, can be obtained from 
\eqref{solh}. In the case considered, ${\cG} = 0$ and $Q = (m-n)/2c$. 

For the massless spinor field, if the nonlinear tern is chosen as
$F = F(J) = \lambda P^\eta$, from \eqref{inv} for $P$ we find 
\begin{equation}
P = D_0/\tau.
\end{equation}
Equation for $\tau$ then takes the form
\begin{equation}
{\ddot \tau} =  \frac{m^2 + n^2}{\tau} +
\frac{\kappa \lambda (\eta -2)D_0^\eta}{2 \tau^{\eta-1}}.
\label{cptau}
\end{equation}
with the solution in quadrature
\begin{equation}
\int \frac{d \tau}{\sqrt{2(m^2 +n^2) {\rm ln}\, \tau 
-\kappa \lambda D_0^\eta\, \tau^{2 - \eta}  
+ E}} = t.
\label{nonlinsolkzp}
\end{equation}
Components of spinor field can be obtained from 
\eqref{solh}. In the case considered, $Q = (m-n)/2c$ and $\Phi = 0$. 

Here we illustrate some numerical results obtained for cases considered above.
The parameters of the equations are taken as follows: for spatial 
inhomogeneity parameters we set $m = 0.0002$ and $n= 0.0001$, whereas 
Einstein's gravitational constant $\kappa$ is taken to be unity.
For the nonlinear spinor field we choose $\lambda = 0.5.$

\begin{figure}
\hspace*{-1.5truecm}
\includegraphics[angle=0,height=7.5cm,width=12.5cm]{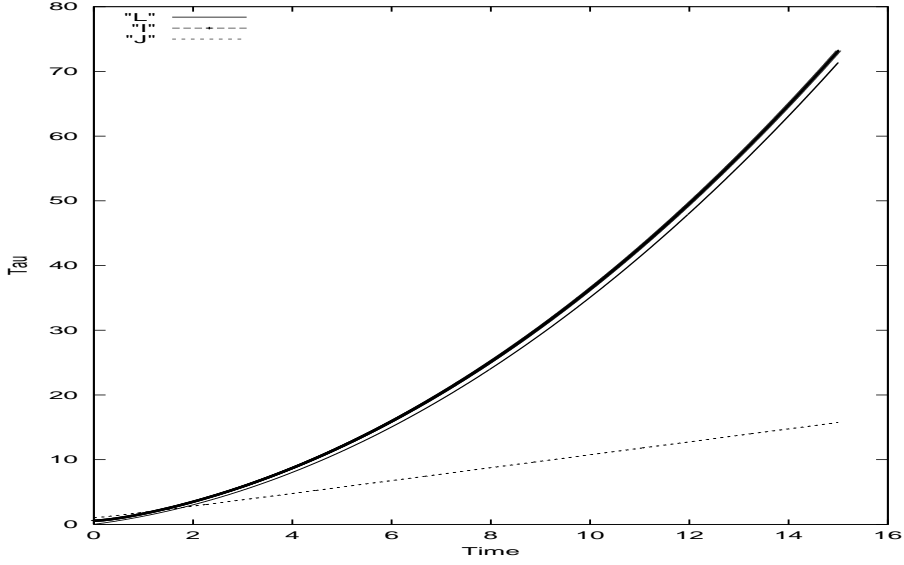}

\caption{Perspective view of $\tau$ for power of nonlinearity
$\eta = 4$ and the integration constant $E = 1$. 
Here ``I'' stands for nonlinear case with $F= F(I)$, ``J'' 
for massless spinor field with the nonlinearity type $F=F(J)$ and ``L'' 
for linear spinor field.}
\label{bvin4}
\end{figure}

As one sees from Fig.~\ref{bvin4}, the presence of the massive term 
accelerates the expansion process. 
For the given parameters in case of linear spinor field, the initial value
of $\tau$ can be chosen very small (here we set $\tau_0= 1.e-6$), whereas
for the nonlinear cases there exists lower boundary, which for the set of 
parameters are $\tau_0 = 0.57$ and $\tau_0 = 1.0$, for ``I'' and
``J'' cases, respectively. Note that for different value of $\eta$ the 
corresponding pictures are similar to Fig.~\ref{bvin4} with the permissible
initial value of $\tau$ increasing with that of $\eta$. Note that the
figures to follow are plotted for the nonlinear term being a function
of invariant $I$, i.e., $F = F(I)$.

\begin{figure}
\hspace*{-1.5truecm}
\includegraphics[angle=0,height=7.5cm,width=12.5cm]{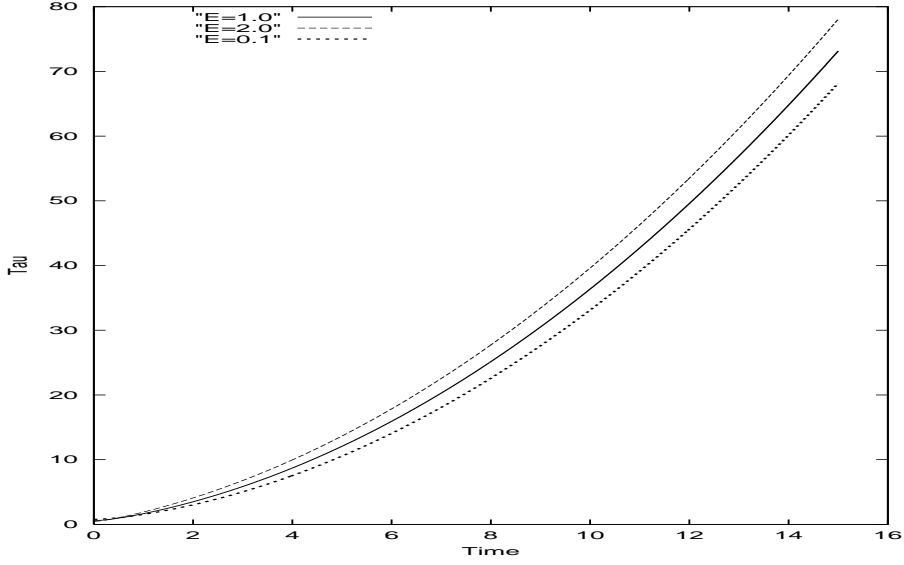}

\caption{Perspective view of $\tau$ for $\eta=4$ and different value
of integration constant $E$.}
\label{bvinE}
\end{figure}
  
As is seen from Fig.~\ref{bvinE}, for a positive $\eta$, the greater is the
value of $E$ the faster the corresponding $\tau$ expands, though
the permissible initial value of $\tau$ for a greater $E$ is less than that
for a smaller one. In this particular case, we have 
$\tau_0 = 0.77,\,0.57,\,0.46$ for $E=0.1,\,1.0,\,2.0$, respectively.
Here we also note that
given the negative value of $\eta$ it is possible to obtain oscillatory
mode of expansion. We do exactly that for the case considered below.

{\bf Case II.} Let us now consider the case setting
\begin{equation}
c = \sqrt{\tau}.
\label{tcsq}
\end{equation}
This leads to following expressions for $a$ and $b$
\begin{equation}
a = \Bigl[\tau^{m/2} {\cal N} {\rm exp}[-\kappa k \int V^0 dt]
\Bigr]^{1/(m+n)},
\label{atoctsq}
\end{equation} 
and 
\begin{equation}
b = \Bigl[\tau^{n/2}/\bigl({\cal N} {\rm exp}[-\kappa k \int V^0 dt]
\bigr)\Bigr]^{1/(m+n)}. 
\label{btoctsq}
\end{equation}

Under this assumption from \eqref{ctau} we get
\begin{equation}
\ddot \tau =  2 \kappa k V^3 \sqrt{\tau} +
 2 (m^2 + n^2) +
\kappa \bigl[MS + {\cD} S + {\cG} P - 2 F\bigr] \tau.
\label{ctausq}
\end{equation}
This equation can be solved exactly as in previous case if we set $k = 0$
and choose the spinor field nonlinearity as $F=F(I)$ or in case
of massless spinor field $F=F(J)$ or $F=F(I\pm J)$.

For the reason that will be given afterwards, we consider the nonlinear 
spinor field in a Bianchi type V (BV) universe setting $m = n$ in the
corresponding equations. To begin with we write the equations for bilinear
spinor forms. Setting $m = n$ in \eqref{inv} one finds
\begin{subequations}
\label{inv5}
\begin{eqnarray}                                
\dot S_0 - 2 {\cG} A_{0}^{0} &=& 0, \\
\dot P_0 - 2 \Phi A_{0}^{0} &=& 0, \\                
\dot A_{0}^{0} + 2 \Phi P_0 + 2 {\cG} S_0 &=& 0,\\
\dot A_{0}^{3}  &=& 0, \\                 
\dot V_{0}^{0}  &=& 0, \\                
\dot V_{0}^{3} + 2 \Phi Q_{0}^{30} - 2 {\cG} Q_{0}^{21} &=& 0,\\
\dot Q_{0}^{30} - 2 \Phi V_{0}^{3} &=& 0, \\                 
\dot Q_{0}^{21} + 2 {\cG} V_{0}^{3} &=& 0,                 
\end{eqnarray}
\end{subequations}
with the following relations between spinor bilinear forms
\begin{subequations}
\label{rel5}
\begin{eqnarray}
(S_{0})^{2} + (P_{0})^{2} + (A_{0}^{0})^2 &=& B_1,\\
A_{0}^{3} &=& B_2,\\
V_{0}^{0} &=& B_3,\\
(V_{0}^{3})^2 + (Q_{0}^{30})^2 + (Q_{0}^{21})^2 &=& B_4,
\end{eqnarray}
\end{subequations}
where $B_i$ are the constant of integration.

Let us now go back to Einstein's equations.
Eqn. \eqref{03} in this case takes the form
\begin{equation}
\frac{\dot a}{a} - \frac{\dot b}{b} = -\frac{\kappa k}{m} V^0.
\label{03v}
\end{equation}
Unlike the BVI, where the corresponding equation, i.e., \eqref{03}
inter-connects all the three metric functions $a,\, b,\, c$, 
Eqn. \eqref{03v} relates only $a$ and $b$ between them:
\begin{equation}
a = {\cal N} {\rm exp}[-(\kappa k/m) \int V^0 dt] b. 
\label{ab2}
\end{equation}
Recalling $\tau = abc$ in view of \eqref{ab2} we can now express $a$ and
$b$ in terms of $\tau$ and $c$:
\begin{eqnarray}
a &=& {\cal N}^{1/2} \sqrt{\tau/c}\,\, {\rm exp}[-(\kappa k/2m)\int V^0 dt], 
\label{atocv}  \\
b &=& {\cal N}^{-1/2} \sqrt{\tau/c}\,\,{\rm exp}[(\kappa k/2m)\int V^0 dt].
\label{btocv}
\end{eqnarray}
In view of \eqref{atocv}, \eqref{btocv} and the fact that $\dot V_0^0 = 0$,
from
\begin{equation}
\frac{d}{d t} 
\Bigl[\tau \frac{d}{dt}\Bigl\{{\rm ln}\Bigl(\frac{b}{c}\Bigr)
\Bigr\}\Bigr]
= -\frac{2m^2}{c^2} \tau,
\label{ab3}
\end{equation}
one obtains
\begin{equation}
\frac{\ddot \tau}{\tau} = 3 \frac{\dot \tau}{\tau}\frac{\dot c}{c}
+ 3 \Bigl(\frac{\ddot c}{c} - \frac{{\dot c}^2}{c^2}\Bigr) - 
4 \frac{m^2}{c^2}.
\label{prom}
\end{equation}
On the other hand \eqref{detertau} in this case has the form 
\begin{equation}
\frac{\ddot \tau}{\tau} = 2 \frac{m^2}{c^2} + \frac{\kappa}{2}
\bigl[3(MS + {\cD} S + {\cG} P - 2 F)\bigr] + \frac{\kappa k}{c} V^3.
\label{detertau1}
\end{equation}
Combining \eqref{prom} and \eqref{detertau1} we obtain 
\begin{equation}
\frac{\ddot c}{c} - \frac{{\dot c}^2}{c^2} + 
\frac{\dot \tau}{\tau} \frac{\dot c}{c} = \frac{\kappa k}{3 c} V^3
+ 2\frac{m^2}{c^2} +
\frac{\kappa}{2} \bigl[MS + {\cD} S + {\cG} P - 2 F\bigr].
\label{ctauv}
\end{equation}
Thus we see that a strait-forward insertion of $m = n$ into 
\eqref{ctau} does not lead to \eqref{ctauv}, since Eqn. \eqref{03}
for different Bianchi type space-time gives different relations between
the metric functions. Here we simply note that for a BIII metric, where
$n = 0$, Eqn. \eqref{03} relates $a$ with $c$, whereas for a BI
universe, as well as for FRW universe there is no such equation.
Note that, though in a BV space-time where $m = n$, many equations in 
question become significantly simpler, it is not enough to write the 
solutions explicitly, since $V^3$, $S$ and $P$ are still undefined 
explicitly. As in previous case, we again consider only time dependent
spinor filed setting $k = 0$. It will give us enough ground to solve
both spinor and gravitational field equations explicitly.
   
Before studying the Eqn. \eqref{ctauv} in details, we go back to 
nonlinear spinor field equations. With $m = n$ and $k = 0$ 
for the spinor field we immediately find
\begin{subequations}
\label{u5}
\begin{eqnarray} 
\dot{u}_{1} + i \Phi u_{1} - {\cG} u_{3} &=& 0, \\
\dot{u}_{2} + i \Phi u_{2} - {\cG} u_{4} &=& 0, \\
\dot{u}_{3} - i \Phi u_{3} + {\cG} u_{1} &=& 0, \\
\dot{u}_{4} - i \Phi u_{4} + {\cG} u_{2} &=& 0. 
\end{eqnarray} 
\end{subequations}

As in BVI space-time we consider the nonlinear term to be $F = F(I)$, or
for massless spinor field $F = F(J)$ or $F = F(I\pm J)$.   
The spinor field Eqn. \eqref{u5} completely coincides with those 
for a BI metric. So in what follows we simply
write the corresponding results without any details. A detailed analysis
of these results can be found in \cite{saprd}.

Thus, for the nonlinear term in the Lagrangian given as $F = F(I)$, 
the components of the spinor field take the form 
~\cite{saprd}
\begin{subequations}
\label{psinl}
\begin{eqnarray} 
\psi_1(t) &=& (C_1/\sqrt{\tau}){\rm exp}\,[-i\int(M - {\cD}) dt],\\
\psi_2(t) &=& (C_2/\sqrt{\tau}){\rm exp}\,[-i\int(M - {\cD}) dt],\\
\psi_3(t) &=& (C_3/\sqrt{\tau}) {\rm exp}\,[i\int(M - {\cD}) dt],\\
\psi_4(t) &=& (C_4/\sqrt{\tau}) {\rm exp}\,[i\int(M - {\cD}) dt].
\end{eqnarray} 
\end{subequations}
Here $C_1,\,C_2,\,C_3,\,C_4$ are the integration constants, 
such that 
$$C_{1}^{2} + C_{2}^{2} - C_{3}^{2} - C_{4}^{2} = C_0,$$
with $ C_0 = S \tau.$

In case, the nonlinear term is given by $F = F(J)$, the components
of the spinor field have the form
\begin{subequations}
\label{psij}
\begin{eqnarray}
\psi_1 &=&\frac{1}{\sqrt{\tau}} \bigl(D_1 e^{i \sigma} + 
iD_3 e^{-i\sigma}\bigr), \\
\psi_2 &=&\frac{1}{\sqrt{\tau}} \bigl(D_2 e^{i \sigma} + 
iD_4 e^{-i\sigma}\bigr),  \\
\psi_3 &=&\frac{1}{\sqrt{\tau}} \bigl(iD_1 e^{i \sigma} + 
D_3 e^{-i \sigma}\bigr), \\
\psi_4 &=&\frac{1}{\sqrt{\tau}} \bigl(iD_2 e^{i \sigma} + 
D_4 e^{-i\sigma}\bigr).
\end{eqnarray} 
\end{subequations}
Here $\sigma = \int {\cG} dt$, and the integration constants
$D_i$ obey 
$$2\,(D_{1}^{2} + D_{2}^{2} - D_{3}^{2} -D_{4}^{2}) = D_0,$$
with $D_0$ to be determined from $ P = D_0/\tau.$ 

Thus we see that in the cases considered here the spinor bilinear forms
are inverse proportional to $\tau$, i.e., $S = C_0/\tau$ and $P = D_0/\tau.$

Let us now go back to \eqref{ctauv}. As one sees, for $k=0$, the assumption
$\tau = c$ makes no sense, since in this case the metric
functions $a$ and $b$ turns out to be constant. So as was mentioned earlier,
we consider the case with $c = \sqrt{\tau}$. Under this assumption from 
\eqref{ctauv} we get
\begin{equation}
\ddot \tau =  4 m^2  +
\kappa \bigl[MS + {\cD} S + {\cG} P - 2 F\bigr] \tau.
\label{ctausqv}
\end{equation}
We consider the case with $F = \lambda S^\eta$. Taking into account that
$S = C_0/\tau$ and ${\cG} = 0$, from \eqref{ctausqv} one derives
\begin{equation}
\ddot \tau = 4 m^2  +
\kappa \bigl[M C_0 + \lambda C_0^\eta (\eta -2)\tau^{1-\eta}\bigr],
\label{ctausqv2}
\end{equation}
with the solution in quadrature
\begin{equation}
\frac{d \tau}{\sqrt{(8 m^2 + 2 \kappa M C_0) \tau - 2 \lambda \kappa C_0^\eta
\tau^{2 - \eta} + E}} = t.
\label{bvqdr}
\end{equation}
Note that, for a linear spinor field and for the massless spinor field 
with $F = \lambda P^\eta$ one has to put $\lambda = 0$ and $M = 0$, 
respectively, into \eqref{bvqdr}. It should be noted that
for a positive constant $E$, in case of linear spinor field $\tau$ may
have even a trivial initial value.

Finally we give numerical solutions in graphical form for case II. It should 
me mentioned that for a positive $\eta$ we have the picture that we obtained
in case I with the expansion being faster in this case. So we concentrate
our study for the negative $\eta$ only. In this case we leave the value
of inhomogeneity parameters and Einstein constant unaltered.

\begin{figure}
\hspace*{-1.5truecm}
\includegraphics[angle=0,height=7.5cm,width=12.5cm]{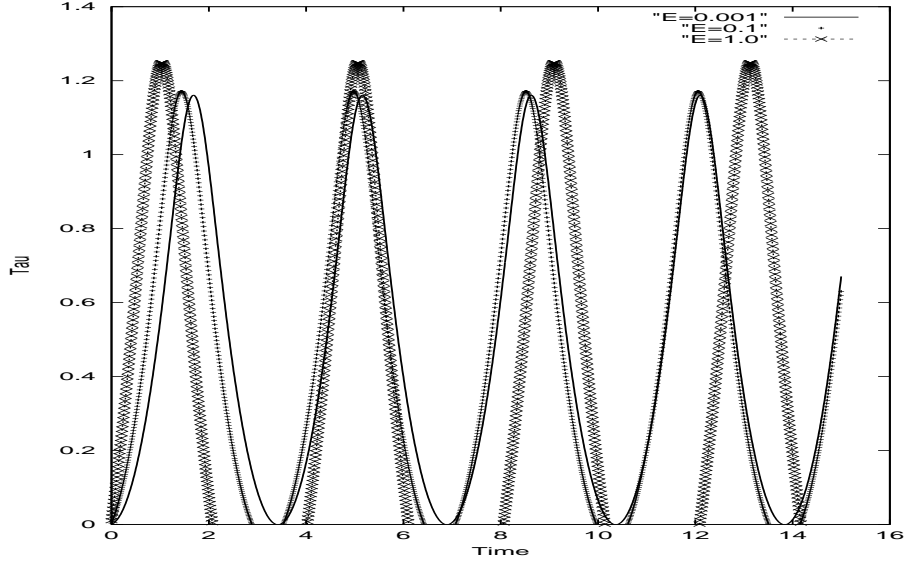}
\caption{Perspective view of $\tau$ for a negative $\eta$ with different
$E$. In this particular case we set $\eta = -4.$}
\label{bvn-4E}
\end{figure}

As one sees from Fig.~\ref{bvn-4E}, a negative $\eta$ gives rise to the
oscillatory mode of evolution. For the given set of parameters, for 
$E \le 0.001$ we have the solution that is almost periodical, i.e., in 
this case the model begins to expand as soon as it reaches the singular
point corresponding $\tau = 0$, whereas for greater $E$ these phases
are separated by a longer time interval.

\begin{figure}
\hspace*{-1.5truecm}
\includegraphics[angle=0,height=7.5cm,width=12.5cm]{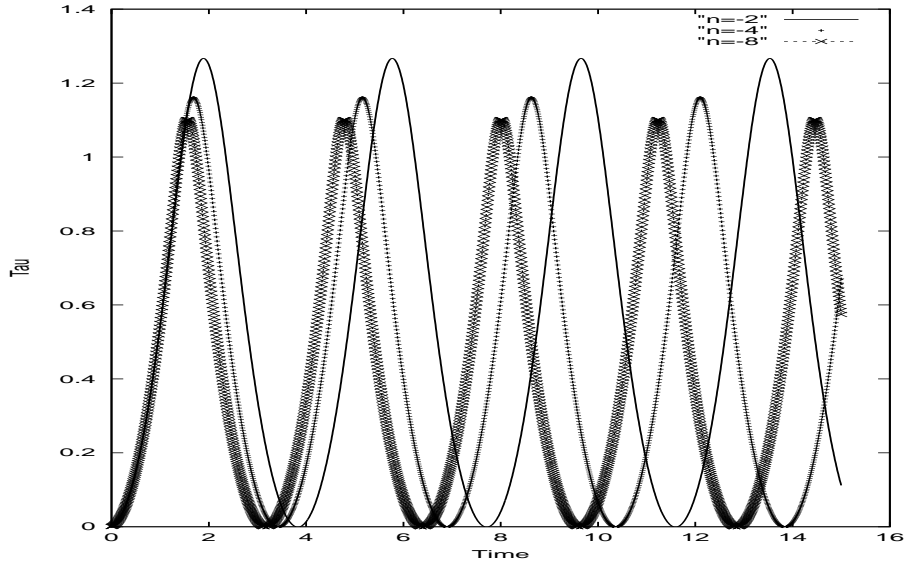}
\caption{Perspective view of $\tau$ for $E =0.001$ with different value of
$\eta.$}
\label{bvEn}
\end{figure}

In figure Fig.~\ref{bvEn} we plot the evolution of $\tau$ for different 
$\eta$ with $E$ being fixed. It should be mentioned that unlike the case
with positive $\eta$ in this case solutions are available only for 
even $\eta$. Finally we like to state that choosing a negative $\eta$
we come to the analogical conclusion for case I.

\section{Conclusion}

Self-consistent system of nonlinear spinor field and gravitational one,
modeled by a Bianchi type VI (BVI) space-time, is studied. Exact solutions 
to spinor and gravitational field equations are obtained for some special
choice of spinor field nonlinearity and space-time inhomogeneity. It is 
shown that if the nonlinear spinor term is chosen to be a function of
the invariants $I=S^2$ or $J=P^2$, with a negative power, the model
provides oscillatory mode of expansion.  
It is shown that though a suitable choice of $m$ and $n$ 
in a BVI metric evokes other Bianchi models, namely, BV, BIII and BI,
the solutions of Einstein equations in these universes cannot be obtained
by simply setting $m$ and $n$ in the corresponding solutions obtained in a 
BVI universe, since in different models the metric functions are 
inter-related differently. Indeed, it follows from the Eqn. \eqref{03},
\begin{equation}
m \frac{\dot a}{a} - n \frac{\dot b}{b}
- (m - n) \frac{\dot c}{c} = \kappa T_{3}^{0}. \label{relation}
\end{equation}
that for a BVI model the metric functions $a,\,b,\,c$ are connected
with each other by \eqref{abcrel}, whereas, for BV universe 
\eqref{relation} gives relation between $a$ and $b$ by \eqref{ab2}
and for a BIII space-time it connects $a$ and $c$. For BI or FRW models
Eqn. \eqref{relation} does not exist.

\end{document}